\documentclass[sigconf,bookmarksnumbered,unicode,hidelinks]{acmart}
\acmConference[SE 2030]{International Workshop on Software Engineering in 2030}{November 2024}{Puerto Galinàs (Brazil)}

\usepackage{graphicx} 
\usepackage{mdframed}
\usepackage{float}
\usepackage{stackengine}
\usepackage{xcolor}
\usepackage{balance}
\definecolor{lightestgrey}{rgb}{0.9,0.9,0.9}
\usepackage{graphicx}
\usepackage{enumitem}
\usepackage{hyperref}
\usepackage{slantsc}
\usepackage{subcaption}
\usepackage{listings}

\newcommand{\etal}[0]{et~al{.}}

\providecommand{\keywords}[1]
{
  \small	
  \textbf{\textit{Keywords---}} #1
}

\begin{document}

\title{Saltzer \& Schroeder for 2030: Security engineering principles in a world of AI}

\author{Nikhil Patnaik}
\email{nikhil.patnaik@bristol.ac.uk}
\affiliation{%
  \institution{University of Bristol}
  \city{Bristol}
  \country{UK}
}

\author{Joseph Hallett}
\email{joseph.hallett@bristol.ac.uk}
\affiliation{%
  \institution{University of Bristol}
  \city{Bristol}
  \country{UK}
}

\author{Awais Rashid}
\email{awais.rashid@bristol.ac.uk}
\affiliation{%
  \institution{University of Bristol}
  \city{Bristol}
  \country{UK}
}

\begin{abstract}

Writing secure code is challenging and so it is expected that, following the release of code-generative AI tools, such as ChatGPT and GitHub Copilot, developers will use these tools to perform security tasks and use security APIs. However, is the code generated by ChatGPT secure? How would the everyday software or security engineer be able to tell?

As we approach the next decade we expect a greater adoption of code-generative AI tools and to see developers use them to write secure code. In preparation for this, we need to ensure \emph{security-by-design}. In this paper, we look back in time to Saltzer \& Schroeder's security design principles as they will need to evolve and adapt to the challenges that come with a world of AI-generated code.

\end{abstract}


\maketitle

\section{Introduction}

In 1974, Saltzer addressed the challenge of protection and control of information sharing in Multics (Multiplexed Information and Computing Service)~\cite{saltzer1974protection}. Saltzer offered 5 design principles to help evaluate different designs. These design principles addressed access control lists, identification and authentication of users, hierarchical control of access specifications, and primary memory protection~\cite{saltzer1974protection}. In 1975, Saltzer \& Schroeder presented 8 design principles and a series of desired functions with the intention of protecting computer-stored information from unauthorized access and modification. At the time software application could store information and be simultaneously used by several users. The key challenge Saltzer \& Schroeder wanted to address was that of multiple use. For applications with users who do not have equal authority, a system is needed to enforce the desired authority structure in the application~\cite{saltzer1975protection}. Saltzer \& Schroeder's work became very influential and applicable to a wide range of fields such as enforcing security policies~\cite{schneider1999enforceable}, evaluating the Java security architecture~\cite{Gong:2003:IJP:599797}, and minimizing user-related faults in information systems security~\cite{siponen2000critical}. However, it wasn't until in 1995 when Saltzer \& Schroeder's principles were first applied to the design of security APIs through Cryptlib~\cite{gutmann1995cryptlib}. Between 1995 and 2002, Gutmann adapted the work of Saltzer \& Schroeder to address the security and usability challenges of designing a cryptographic API~\cite{gutmann1995cryptlib,gutmann1999design,gutmann2002cryptosoftware}. Gutmann's work played an integral role in introducing Saltzer \& Schroeder's design principles to the field of security API design~\cite{green2016developers,patnaik2019usability,acar2017comparing,oliveira2018APIblindspots,votipka2020securitymistakes,mindermann2018rust}. The evolution of Saltzer \& Schroeder's principles can be accredited to the ever-changing landscape of cyber security and the challenges that come with it as it is these challenges that force principles to evolve.
 
Today, we are witnessing the advent of an AI world. With the release of large-language models (LLMs), OpenAI began an AI race with an undefined finish line. Now running against competitors such as Google's PaLM and LaMDA, we are seeing rapid developments in this field. The software development community has shown great interest in these advancements as the integration of code-generative AI tools such as ChatGPT and GitHub Copilot may help them in many diverse tasks such as improving the code review process, providing coding suggestions, suggesting fixes for defects in code. In the future, these code-generative AI tools could help in refactoring the code bases of large, complex software applications, improving readability and making it easier to maintain with the support of AI-generated documentation~\cite{bird2022taking,ernst2022ai}.

With all these upcoming changes in the software industry, what role do Saltzer \& Schroeder play in this AI world? Are their security design principles enough? If not, why not? If they are enough, how should we use them to shape the vision for 2030.

In this paper, we evaluate ChatGPT's ability to securely store a password and assess the generated code against Naiakshina's criteria for secure password storage~\cite{naiakshina2017developers}. Following this, we prompt ChatGPT to repeat the task but this time using Naiakshina's security criteria as a list of requirements. Finally, we ask ChatGPT to assess its generated code against the security design principles of Saltzer \& Schroeder~\cite{saltzer1975protection}. The purpose of this exercise is not only to understand ChatGPT's capabilities in secure coding and understanding of security principles but also as a step in understanding where in the Developer/AI dynamic can Saltzer \& Schroeder be introduced? Where in the software development process would Saltzer \& Schroeder be most effective? If they are not effective, what are the shortcomings? Can they be addressed through an adapted Saltzer \& Schroeder set of principles? Are Saltzer \& Schroeder enough or is it the end of the road?

\section{Background}

\subsection{General insights}

Following the release of GitHub Copilot, the AI pair-programmer, Bird~\etal~\cite{bird2022taking} studied the process of pair programming and how the integration of GitHub Copilot changed the role of the developer. Bird~\etal{} explains how when two developers write code together, one takes the role of the \emph{driver}, writing the code for the assigned task, while the other works as the \emph{navigator}, directing and reviewing the driver's work. However, Bird~\etal{} found that once GitHub Copilot is introduced, the developers role shifts from \emph{writing} the code to \emph{understanding} the code. The introduction of GitHub Copilot means that the skill of reviewing code has become equally, if not more important than writing code. The means by which developers methods to assess AI-generated code are yet to be developed. As for now, the generated code undergoes the same review process as a developer~\cite{bird2022taking, ernst2022ai}.


In terms of productivity, GitHub Copilot has been known to help developers adapt to different languages and writing styles~\cite{bird2022taking}. However, in regards to code quality, Yetiştiren~\etal{}~\cite{yeticstiren2023evaluating, yetistiren2022assessing} performed an empirical study to compare the code quality and correctness of GitHub Copilot, ChatGPT, and Amazon CodeWhisperer using the HumanEval dataset~\cite{chen2021evaluating} and found that, in terms of code correctness, ChatGPT performed the best with correct solutions for 65.2\% of the HumanEval dataset problems, followed by GitHub Copilot. Yetiştiren did note developers should expect to find some bugs with all these tools but that these bugs are not as common as code smells identified. There were a total of 13 code smells identified from the three code-generative AI tools~\cite{yeticstiren2023evaluating, yetistiren2022assessing}.

Ernst~\etal{}~\cite{ernst2022ai} comment on the short and long-term effects of using AIDEs (Artificial Intelligence Driven Development Environments) for learning how to code. On one hand tools like ChatGPT and GitHub Copilot can really help a novice developer learn a programming language by offering suggestions and helping with coding tasks. This might seem appealing to many developers joining industry today but on the other hand these developers may struggle to fully understand why a suggested piece of code works and how to correctly integrate it with their existing code base~\cite{ernst2022ai}. Furthermore OpenAI's Codex poses a challenge to lecturers and university modules as it is good enough to surpass many students in `Computer Science 1' modules of a Computer Science degree. Students may begin to think that if Codex is better at coding than they are, why should they bother learning how to code? Instead, students will start focusing on prompt engineering and learning how to fine-tune AI-generated code. Their role shifts from driver (the code writer) to a navigator (a task designer) but this comes with serious consequences. Many students are first introduced to coding practices and design principles at university and so by introducing LLMs like Codex, students may be robbed of applying design principles they learn in theory to practical applications as they may ask ChatGPT to generate code for them and will lose their ability to problem solve and think critically. This potential scenario also stops the adoption and evolution of security design principles and secure coding practices~\cite{saltzer1975protection, gamma1993design, nielsen1994enhancing, fowler2018refactoring, bloch2001effective}.

In a recent study by Lau~\etal{}~\cite{lau2023ban}, a survery of 20 programming instructors was conducted to understand how they plan on adapting to the introduction of AI models. For the short-term, practitioners intend to discourage the use of such tools but in preparation for the future, some agreed that such resistance is futile and are willing to integrate code-generative AI tools with their existing program.

\subsection{Security implications of using AI models}

Recent works have taken steps to understand the security implications of using code-generative AIs tools~\cite{he2023large,hajipour2023systematically, siddiq2022securityeval, tony2023llmseceval,pearce2022asleep}. Pearce~\etal{}~\cite{pearce2022asleep} ran an empirical study to assess the GitHub Copilot's performance against a subset of MITRE's Top 25 Common Weakness Enumerations (CWEs). The results, from a security standpoint, were indefinitive but Pearce~\etal{} note that GitHub Copilot is trained on an open-source dataset and so bugs should be expected by developers and so they are advised to remain critical and vigilant of the code generated by GitHub Copilot. An interesting point mentioned by Pearce~\etal{} is the effect of time on the quality of security in open-source software. The best practice at the time of writing may slowly become bad practice as the cyber security landscape changes.

He~\etal{} devise a novel security task called \emph{controlled code generation} for which they propose SVEN a learning-based approach model with an additional binary parameter added alongside the prompt. Through this binary parameter He~\etal{} can specify the process of security hardening or adversarial testing. Similarly to Pearce~\etal{}~\cite{pearce2022asleep}, SVEN is evaluated using a subset of MITRE's Top 25 CWEs. He~\etal{} found that SVEN achieves a strong level of security control through their novel task.

Hajipour~\etal{}~\cite{hajipour2023systematically} address the challenge of finding security vulnerabilities in the Codex and CodeGen LLMs using an inversion few-shot approach. In this approach, AI models are prompted using examples of vulnerable code and supporting prompts to see whether it generates vulnerable code. As with all the studies discussed here, Hajipour~\etal{}~\cite{hajipour2023systematically} use CodeQL to evaluate the security of the generated code. As a result of their study, they showed that their approach is capable of finding thousands of security vulnerabilities in the AI models but more works needs to be done to define a better methodology for identifying prompts that lead to security vulnerabilities~\cite{hajipour2023systematically}. Tony~\etal{}~\cite{tony2023llmseceval} actually used vulnerable code snippets to their advantage to help evaluate the GPT-3 and Code LLMs ability to generate secure code. They presented LLMSecEval a dataset of 150 natural language prompts with descriptions of code snippets prone to vulnerabilities seen in MITRE's Top 25 Common Weakness Enumerations (CWEs). Each prompt from the dataset had a secure implementation associated with it to compare GPT-3 and Codex responses. A similar approach can be taken to evaluate the presence of security design principles in AI-generated code.



\section{Case Study: Secure Password Storage with ChatGPT}

We present the case study of \emph{securely storing a password}, a well-known security task that has been studied extensively by the academic research community~\cite{hallett2021and,naiakshina2017developers,he2023large}. However, the literature shows that developers have issues with the technicalities of securely storing passwords~\cite{hallett2021and,naiakshina2017developers,he2023large}. Furthermore, there is well-established criteria in literature by Naiakshina~\etal{}~\cite{naiakshina2017developers} to evaluate the security of such password storage. Hence this serves as a useful case study to analyze security of ChatGPT's responses.

\subsection{ChatGPT Without Explicit Security Requirements}

When plainly asked to store a password securely across 5 prompt windows, ChatGPT offers implementations that vary in terms of libraries used: \emph{bcrypt}, \emph{getpass}, and \emph{hashlib}. We measured the implementations' level of security against Naiakshina's criteria for securely storing a password~\cite{naiakshina2017developers}.

\begin{figure} [!ht] \centering
  \begin{mdframed}\raggedright
    \begin{itemize}[leftmargin=*]
      \item The end-user password is salted (+1) and hashed (+1).
      \item The derived length of the hash is at least 160 bits long (+1).
      \item The iteration count for key stretching is at least 1,000 (+0.5) or 10,000(+1) for PBKDF2 and at least $2^{10}$ for bcrypt (+1).
      \item A memory-hard hashing function is used (+1).
      \item The salt value is generated randomly (+1).
      \item The salt is at least 32 bits in length (+1).
    \end{itemize}
    \end{mdframed}
  \caption{Naiakshina's end-user password storage assessment
criteria~\cite{naiakshina2017developers}, copied verbatim.  A score
$\mathbf{\geq 6}$ indicates industrial best practice.}
  \label{fig:naiakshina}
\end{figure}

The implementations scored an average of 3 of out 8 according to Naiakshina's criteria~\cite{naiakshina2017developers}. Four out of the five implementations used the \emph{bcrypt} library for the task followed by the common use of the hashpw() and gensalt() functions to hash and salt the password. This gained the implementation the first two points available from the criteria. The third point came from the salt being 32 bits long. Steps such as key stretching, use of memory hard hash functions and hash lengths being at least 160 bits were not addressed in ChatGPT's responses. 


\subsection{ChatGPT with Explicit Security Requirements}

We asked ChatGPT to write code that securely stores a password using Naiakshina's criteria as a list of security requirements. The five implementations generated scored an average of 5 out of 8, with one of the implementations scoring 8 out of 8. These five implementations implemented cryptographic algorithms and features such as \emph{PBKDF2}, \emph{key stretching}, and using other hashing algorithms such as \emph{SHA-256}. We did not see recommended memory-hard hashing functions such \emph{scrypt} and \emph{Argon2d}. In this short test, we see that by defining a security criteria, ChatGPT offers responses that are more secure. The C Secure Coding Standard contains hundreds of non-compliant code examples, which served as a ready-made dataset of coding errors Sherman~\cite{Sherman2024ChatGPTAnalyze} ran through ChatGPT 3.5, and evaluated the responses against the patched code provided. When it comes to analyzing code, ChatGPT correctly identified the problem 46.2\% of the time. This meant that 52.1\% of the time, ChatGPT did not identify the coding error. The remaining 1.7\% is attributed to aesthetic-related issues rather than an error. ChatGPT showed promise but the limitations were clear~\cite{Sherman2024ChatGPTAnalyze}. Sherman noted that the underlying mechanism used by LLMs depends on pattern matching based on the training data. We found that the  short-comings of ChatGPT's program analysis skill can be mitigated by prompt engineering. By fine-tuning prompts and specifying security criteria, ChatGPT generates more secure code.

Building on Bird~\etal{}'s \emph{driver/navigator} model, we can say that although the introduction of code-generative AI tools like ChatGPT encourages the shift of the developers role from a driver (the code writer) to a navigator (a task designer), the encouraged shift may be premature~\cite{bird2022taking}. It is well documented that developers, in their current role as a driver, find it challenging to write secure code themselves~\cite{nadi2016jumping, georgiev2012mostdangerouscode, acar2017comparing}. The challenges can be attributed to usability issues surrounding security APIs and simply having a lack of security expertise to navigate these APIs. What is important to note here is that software developers would take the initial implementations, written without Naiakshina's security criteria at face value and consider it secure. This makes them not only unfit as the driver but also unprepared for taking on the role of the navigator. The role of a navigator requires the developer to design the task and the security requirements to achieve it. Without security principles to help developers assess the security of AI-generated code, developers' applications are open to exploitation and vulnerabilities~\cite{siddiq2022securityeval,tony2023llmseceval}.

\subsection{ChatGPT and Saltzer \& Schroeder}

While OpenAI's GPT model does not inherently follow the design principles of Saltzer \& Schroeder, it is capable of remembering the context of a conversation. This meant that once we brought up the Saltzer \& Schroeder design principles, the model tried its best to adhere to their principles whenever it was asked to generate code. ChatGPT also explained, in each response, how the code it generated either followed Saltzer \& Schroeder or did not. To better understand the process, we have shared the principles addressed by ChatGPT's response and our evaluation below:

\begin{itemize}
  \item \textbf{Least Privilege}: ChatGPT: ``We can see that the code limits its access to only the necessary functionality needed to achieve the task''. However, while successfully adhering to the \emph{least privilege} principle, ChatGPT has shown the saltCount parameter of the getSalt() function as unnecessary when, in fact, it is necessary to strengthen the password storage~\cite{naiakshina2017developers}.
  
  \item \textbf{Fail-safe Defaults}: As we noted in Section 3.1, 4 out of 5 implementations from ChatGPT used the bcrypt library. When it comes to using a hashing algorithm, bcrypt use the Blowfish algorithm as a fail-safe default which is acceptable but there are more secure algorithms available. 
  
  \item \textbf{Complete Mediation}: As we mentioned before, Saltzer \& Schroeder's principles have influenced the works of others through adaptations over time. ChatGPT: ``All the functions in the code generated by ChatGPT are mediated through the bcrypt library. Developers do not need to implement their own cryptographic functions or go looking elsewhere, reducing the risk of implementation errors.''

  However, complete mediation requires trust and as we saw, against Naiakshina's security criteria, ChatGPT does not perform well if the criteria is not defined in the prompt. This is problematic as the everyday software developer will not know to add security criteria as part of their prompt and so will take the generated code on face-value and consider it secure even if it is not. This problem of weak security can be easily identified in a controlled case study like this one but what are the identifiers and mitigations for software and security engineers in the real world, an AI world in 2030? How do we ensure that the security of AI-generated code is up to standard? Is this where Saltzer \& Schoreder come in?
  
  \item \textbf{Psychological Acceptability}: Designed with the everyday developer in mind, the bcrypt library offers a straightforward cryptographic API for securely storing a password without developers needing a strong level of cryptographic expertise, achieving psychological acceptability. However, a quick search through StackOverflow shows that there are quite a few questions associated with the bcrypt library, many of which are issues to do with building the library, trouble installing it, difficulty using it with other applications such as MySQL, NodeJS, and cross-language hashing. This informal search shows that bcrypt has the \emph{Doesn't play well with others} usability smell~\cite{patnaik2019usability}.
  
  \item \textbf{Separation of Privilege}: The bcrypt library supports the functionality of password hashing and verification, so that developers can enforce separation of privilege by restricting access to the password database and related operations.
\end{itemize}

We can see that ChatGPT is able to apply some of Saltzer \& Schroeder's principles to the task of securely storing a password but its application of these principles shows an incomplete application and introduces a series of trust-related issues as well.

\section{Discussion}

\balance

What role do Saltzer \& Schroeder play in this AI world?

To answer this question lets first look at the challenges Saltzer \& Schroeder initially set out to address and compare them to the challenge we may see from 2030. The key challenge Saltzer \& Schroeder wanted to solve was that of multiple use, multiple user having access to information on one shared computer. The principles they defined were proposed for a system that could enforce the desired authority structure in the application~\cite{saltzer1975protection}.

What is the `multiple use' challenge for security engineering today? It's not a handful of people sharing a computer. Its scaled up, instead its many people, many developers, many security engineers working on securing giant applications and an LLM trained code-generative AI tool is now involved and part of today's `multiple use' issue too. Let's apply some of Saltzer \& Schroeder's principles to OpenAI's GPT model. \emph{Economy of Mechanism} is easily achieved through a simple, easy-to-use interface in ChatGPT. However, we did find that in order to see security in its code, one needs to define security in one's prompt. Unwanted access paths are seen through the poor implementation of securing a password and will not be noticed by the everyday developer. 

By default, ChatGPT does not respond to requests of malicious code but there are work-arounds. When it comes to security engineering and writing secure code, ChatGPT has no \emph{Fail-Safe Defaults} of its own. It relies on the training data to be ready-made secure. Unless prompted~\cite{naiakshina2017developers}, ChatGPT cannot factor in the strength of cryptographic primitives and secure coding practices when writing secure code. Recent works have evaluated LLMs on vulnerability-prone datasets~\cite{pearce2022asleep,tony2023llmseceval,Sherman2024ChatGPTAnalyze, hajipour2023systematically}. In this case study, we attempted to integrate Saltzer \& Schroeder's principles with \emph{Prompt Engineering}, but is it possible to introduce Saltzer \& Schroeder's principles through a different avenue, through secure-design focused training?

The \emph{Open Design} principle states that the mechanism should not depend on the ignorance of potential attackers but the adoption of LLMs through tools like ChatGPT does implicitly encourage the ignorance of its users~\cite{bird2022taking, ernst2022ai}. Bird~\etal{} captured the changing dynamic in the developers role after the introduction of GitHub Copilot. It begins with the automation of mundane tasks being given to AI models, leaving us developers with the more creative tasks but when it comes to reviewing the code generated by Copilot, Bird~\etal{} noted developers showing some hesitancy and feelings of being unsure whether the code is functionally correct or not~\cite{bird2022taking}. To balance this argument, we should mention that novice developers do say that they benefit from using AI models to learn new programming languages~\cite{bird2022taking}.

We found that ChatGPT could only apply some of Saltzer \& Schroeder's principles and for the ones that it did, we pointed out many misapplications of the principles showing a incomplete understanding on ChatGPT's behalf. We argue that Saltzer \& Schroeder do have a role to play in the upcoming decade. With fine-tuning and prompt engineering, its possible that Saltzer \& Schroeder's principles are better applied by LLMs.

To address the issue of a lack of fail-safe-defaults and general security application, Saltzer \& Schroeder's principles could be reworked into more actionable security standards tailored towards a security-based training set for LLMs. Considering the sophisticated types of attacks available and LLMs susceptibility to vulnerable datasets, we don't think Saltzer \& Schroeder's principles are enough as they are and they need to be reworked and adapted to address challenges such as ML-poisoning for example. If we, play out the `multiple use' scenario, ML-poisoning introduces vulnerabilities into LLMs making the LLM itself vulnerable.

Finally, we need to address the consequences of the developers shifting role from a driver (the code writer) to a navigator (a task designer). We think that this shift will continue into the 2030, however, its important to continue supporting developers and encouraging them to use security APIs and tools themselves today. We should not only apply Saltzer \& Schroeder's principles but the many other sets available such as Gamma~\etal{}'s design patterns, Nielsen's usability heuristics~\cite{nielsen1994enhancing}. These principles and guidelines can be applied to improving the design of existing security APIs, like OpenSSL, and the development of new security tools. 

\section{Research Agenda 2030}

The software engineering community needs to drive research forward in:

\begin{itemize}
  \item LLM Fine-tuning: Fine-tune LLMs to better adhere to the Saltzer \& Schroeder principles~\cite{saltzer1975protection}.
  
  \item Security-design training dataset: Focus on developing a security design based training set. 

  \item Security Tools and API Development: Continue helping develpers write secure code with security tools and APIs to feed the `driver'.

Our vision for 2030 is that we'll have a security-focused LLM trained on Saltzer \& Schroeder validated code. With fine-tuning, our LLM can start taking on more advanced security tasks. These developers will have a wide range of security tools and the requisite skills to support in writing secure code.
  
\end{itemize}

\bibliographystyle{ACM-Reference-Format}
\bibliography{bibliography}

\end{document}